\documentstyle[osa,manuscript,graphicx,epsfig]{revtex4}  
\newcommand{\rhz}{$\sqrt{{\rm Hz}}$}

\newcommand{\bega}{\begin{array}}
\def\endar{\end{array}}
\newcommand{\beq}{\begin{equation}}
\newcommand{\eeq}{\end{equation}}
\newcommand{\bef}{\begin{figure}}
\newcommand{\enf}{\end{figure}}
\newcommand{\bec}{\begin{center}}
\newcommand{\enc}{\end{center}}
\newcommand{\bei}{\begin{itemize}}
\newcommand{\eni}{\end{itemize}}
\newcommand{\ingr}{\includegraphics}

\newcommand{\om}{{\omega}}

\newcommand{\ith}{{\em i}-{\small th} }

\begin{document}                
\title{Inertial control of the mirror suspensions of the VIRGO interferometer for gravitational
wave detection.\footnote{{\bf Submitted} to {\it Review of
Scientific Instruments}}}

\author{G.Losurdo\footnote{Corresponding author, e-mail: losurdo@pi.infn.it},
G.Calamai$^1$, E.Cuoco$^1$, L.Fabbroni,\\ G.Guidi$^2$, M.Mazzoni$^1$, R.Stanga$^1$, F.Vetrano$^2$}
\address{Istituto Nazionale di Fisica Nucleare - Sezione di Firenze \\
$^1${\it also}: Universit\`a degli Studi di Firenze \\
$^2${\it also}: Universit\`a degli Studi di Urbino}

\author{L.Holloway\footnote{Now at INFN Pisa}}
\address{University of Illinois, Urbana, IL}

\author{D.Passuello, G.Ballardin, S.Braccini, C.Bradaschia, R.Cavalieri, R.Cecchi,
G.Cella$^3$, V.Dattilo, A.Di Virgilio, F.Fidecaro$^3$, F.Frasconi,
A.Gennai\footnote{Now at ESO, Garching, Germany.}, A.Giazotto,
I.Ferrante,\\ P.La Penna, F.Lelli, T.Lomtadze,
A.Marin\footnote{Now at University of Padova and INFN - Sez. di
Padova.}, S.Mancini, F.Paoletti, A.Pasqualetti, R.Passaquieti$^3$,
R.Poggiani$^3$, R.Taddei, A.Vicer\`e\footnote{Now at Caltech,
Pasadena, California.}, Z.Zhang}
\address{Istituto di Fisica Nucleare - Sez. di Pisa \\
$^3${\it also}: Universit\`a degli Studi di Pisa }

\maketitle
\newpage
\begin{abstract}                
interferometric gravitational wave detectors must be isolated from
seismic noise. The VIRGO vibration isolator, called {\em
superattenuator}, is fully effective at frequencies above 4 Hz.
Nevertheless, the residual motion of the mirror at the mechanical
resonant frequencies of the system are too large for the
interferometer locking system and must be damped. A
multidimensional feedback system, using inertial sensors and
digital processing, has been designed for this purpose. An
experimental procedure for determining the feedback control of the
system has been defined. In this paper a full description of the
system is given and experimental results are presented.

\end{abstract}

\newpage

\section{Introduction}

The sensitivity of interferometric antennas for gravitational wave detection
\cite{ligo,virgo,geo,tama,aigo} is limited at low frequencies by seismic noise. In
order to suppress seismic noise below the thermal noise level
above 4 Hz, a special vibration isolator
has been designed to suspend the mirrors of the VIRGO detector: the superattenuator (SA)
\cite{artsa}.
The expected residual motion of the mirror is $\sim 10^{-18}$ m/\rhz @4
Hz. At lower frequencies, the residual motion of the mirror is much larger ($\sim
0.1$ mm RMS), due to the normal modes of the SA (the resonant frequencies of the
system are in the range 0.04-2 Hz).

In order to maintain the VIRGO interferometer locked \cite{lock} the RMS motion
of the suspended mirrors must not exceed $10^{-12}$ m.
The VIRGO locking strategy is based on a hierarchical control. Feedback forces can
be exerted on three points of the SA: the inverted pendulum (IP) \cite{ip} suspension point,
the marionette (a stage properly designed to steer the
mirror\cite{mario}) and the mirror itself, by means of a seismic noise free recoil mass.
The control on the three points is operated in
different ranges of frequency and amplitude. The maximum mirror displacement that
can be controlled from the marionetta without injecting noise in the detection band
is $\sim 10$ $\mu$m. Moreover, the lower the residual mirror motion the shorter the
time needed for the interferometer locking. Therefore, damping of the SA normal
modes is required for a correct operation of the locking system. A wideband high
gain active control of the SA normal modes using sensors and actuators on top of
the IP and capable of reducing the mirror residual motion within a few microns has
been successfully implemented. It has been defined {\it inertial damping} since it
mostly makes use of accelerometers.

Several previous works have been done in this field. The use of
accelerometers for vibration isolation in gravitational wave experiments
had already been proposed many years ago (see ref. \cite{oldsaulson} and ref.
therein). More recently a multistage active vibration isolation system using
multidimensional active controls has been proposed for the LIGO detector
\cite{jila,stiff}.

\section{Experimental setup}

The experimental setup (fig. \ref{setup}) consists of a full scale SA, provided
with three accelerometers (placed on the top of the IP), three LVDT position
sensors (measuring the relative motion of the IP with respect to an external frame)
and three coil-magnet actuators. The sensors and actuators are all placed in a
pin-wheel configuration. The sensors and actuators signals are processed by a
computer controlled ADC (16 bit)-DSP-DAC (20 bit) system. The DSP handles the
signals of all the sensors and actuators, recombines them by means of matrices,
creates complex feedback filters with high precision pole/zero placements and
performs calculations at a high sampling rate (10 kHz).

In the following we briefly describe the main features of the digital electronics
designed for the VIRGO active controls:
\begin{itemize}
\item {\bf DSP:} the Virgo control system runs at 10 kHz, in order not to have excessive phase
rotation at frequencies of interest (we foresee controlling the suspended masses of
the interferometer up to 100 Hz). With a sampling frequency of 10 kHz, the input to
output delay of our DSP system (taking into account the delay introduced by the
anti-aliasing filter in front of the ADC and the corresponding low pass filter
after the DAC) is about 540 $\mu$sec. This delay will introduce a phase rotation of
about 20 degrees at 100 Hz, low enough for our purposes. On the other hand the
system must be able to deal with very low frequencies (the main resonance of the
inverted pendulum is about 40 mHz) and this requires a very high arithmetic
precision in the DSP system: it can be shown that, for a second order
filter, the minimum frequency that can be implemented is proportional to the square
root of the arithmetic precision. For instance, the minimum frequency for a filter
with a quality factor of 10 (at a sampling frequency of 10 kHz) is about 2.5 Hz
with a mantissa of 24 bits, whereas it is about 0.15 Hz with a mantissa of 32
bits.
The chosen processor was the DSP 96002 from Motorola, that
supported the extended arithmetic precision (32 bits of mantissa) needed for our
purposes.

\item {\bf ADC:} two sources contribute to the noise of an analog to digital converter: the
electronic noise and the quantization noise due to the limited number of bits of
the converter. The ADC is a 16 bit system with an input range of $\pm$10 Volt; the
first noise source slightly dominates the second so that our ADC is equivalent to
an ideal 15 bit ADC. The RMS value of the total noise is about 170 $\mu V$; its
power spectral density is uniformly distributed on the Nyquist band. At a sampling
frequency of 10 kHz this corresponds to a spectral density of about 2.4 $\mu
V/\sqrt{{\rm Hz}}$. In order to lower the noise level it is necessary to increase the
sampling frequency. Our ADC is able to sample the input signal up to 200 kHz and to
send to the DSP the sum of the last N samples, where N ranges from 1 to 256.
Working with the ADC at 200 kHz and N equal to 20, we get a noise level of about
500 $nV/\sqrt{{\rm Hz}}$. This corresponds to a dynamic range of about 143
dB$\cdot\sqrt{{\rm Hz}}$ (we define the dynamic range of a system as the ratio
between the maximum RMS value the system is able to manage and the spectral density
noise level).

\item {DAC:} our DAC system is based on an audio 20 bit
converter: its measured noise level is less than 250 $nV/\sqrt{{\rm Hz}}$.

\end{itemize}

The dynamic range of our digital system is almost equivalent to that of a good
analog one and can be further improved by using pre-emphasis and
de-emphasis filters. In addition a digital system is much more flexible for what
concerns the complexity of the filters that can be implemented, the precision and
the time stability of the filter parameters.

\section{From a MIMO to a SISO system}

The IP is a three degrees of freedom mechanical system: three independent sensors
are required to fully determine its position and three independent actuators to
move the IP in the required settings. The sensors and the actuators are mounted on
the top stage in triangular configuration. Each sensor is, in principle, sensitive
to movements in all the three IP normal modes (which are defined as
$x,y,\theta$, although they do not correspond necessarily to pure translations and
rotations). In the same way, each actuator will generate movements of the IP
involving a mix of the three modes. The basic idea of the IP controls, which
applies both to the inertial damping and the position control, is to {\em
diagonalise} the sensing and control actions: the aim is to  pass from the
sensor/actuator space, to a space where each normal mode is independently sensed
and acted upon. Mathematically, this means to realize a coordinate transformation
such that the equations of motion have the form:
\beq
\ddot{x}_i+\om_i^2 x_i=q_i
\eeq
where the $x_i$ (for $i=1,2,3$) is a normal coordinate,
$\om_i/2\pi$ is the resonant frequency of the \ith mode and $q_i$ is the
generalized force corresponding to the coordinate $x_i$ \cite{goldstein}.
This means to find three linear combinations of the sensor outputs,
defined {\em virtual sensors}, each sensitive to a single normal mode and,
correspondingly, three linear combinations of the excitation coil currents ({\em
virtual actuators}) which excite each mode separately. In control theory
terminology, this means to break down a {\em multiple in-multiple out} (MIMO)
system into many {\em single in-single out} (SISO) systems. The control of a SISO
system is much easier: every mode is controlled by an independent feedback loop,
simplifying greatly the loop design and the stability requirements. In this chapter
we describe a possible approach to the problem. A more general approach would require
the system description in the
state space representation (see for instance ref. \cite{libro,ipcontr}).

\section{Diagonalization: the parameter search scheme.}

Several approaches to the diagonalisation problem have been defined. In this paper
we describe the fastest one. Other procedures are described in ref.
\cite{tesi,ipcontr}.

In the parameter search scheme the sensor/driver/mechanical-response system is
described by a set of parameters. Sets of excitation/response data are taken and
stored on disk.  A merit function is constructed that is related to the degree of
success of diagonalization.  The space of system parameters is then searched using
standard algorithms in order to maximize the merit function.

\subsection{Merit Function}
If the resonance peaks are well separated from each other the value of the
imaginary part of the transfer function of an arbitrary, 'real',  sensor shows two
distinct peaks in the region of the IP translational modes. A diagonalized,
'virtual', sensor should show only one peak. We construct a merit function by
integrating the (absolute) value under the two peaks. If the diagonalization is
correct, then there should be only a small  amount of $y$-mode in the $x$-peak and vice
versa.  In addition, the amount of the rotational mode present in the $x$ and $y$ peaks
should be small. The merit function is constructed using a weighted sum of
appropriate ratios of these integrated values.

\begin{figure}
\begin{center}
\begin{minipage}{.47\textwidth}
\ingr[width=.95\textwidth]{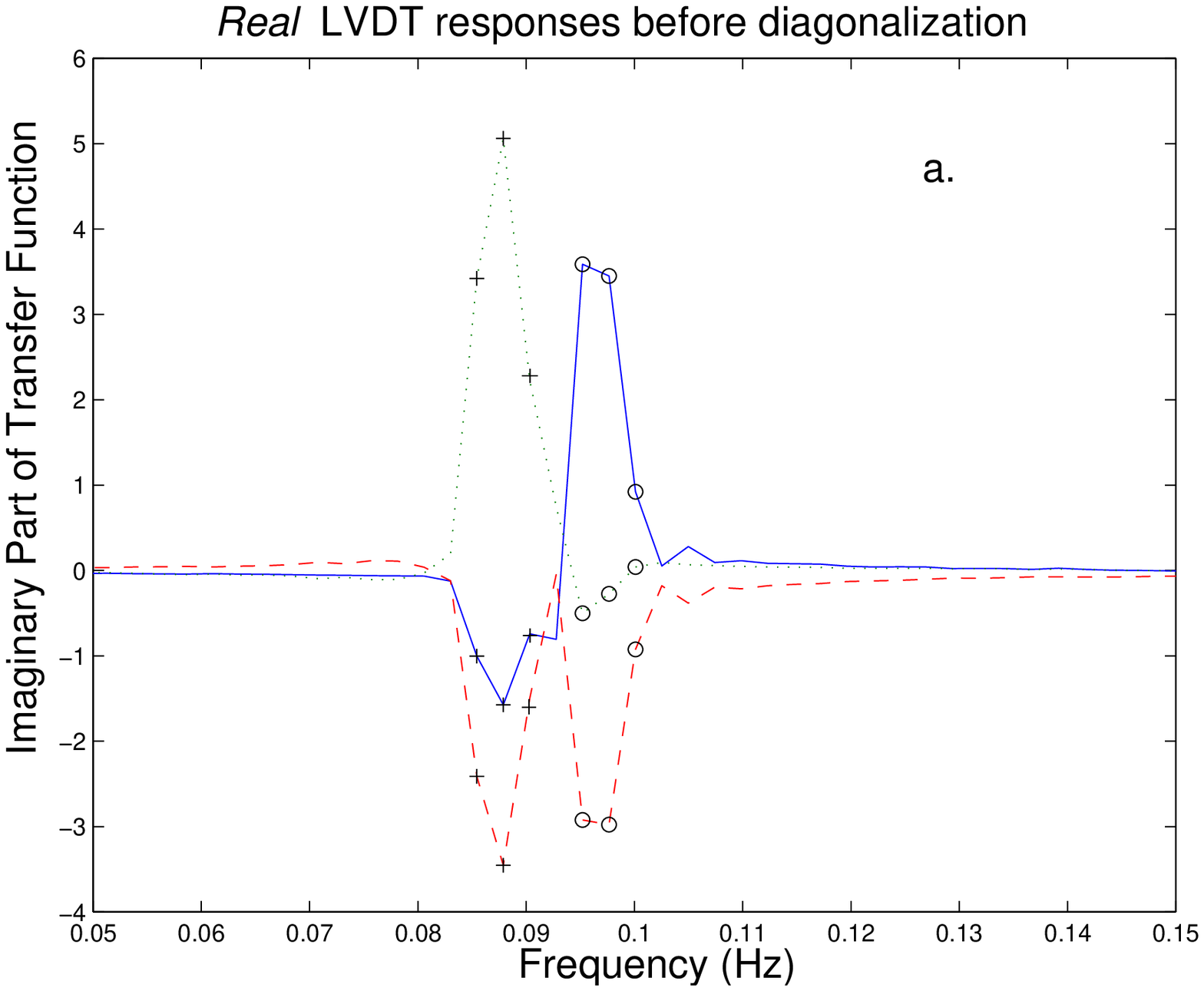}
\end{minipage}\hfill
\begin{minipage}{.47\textwidth}
\ingr[width=.95\textwidth]{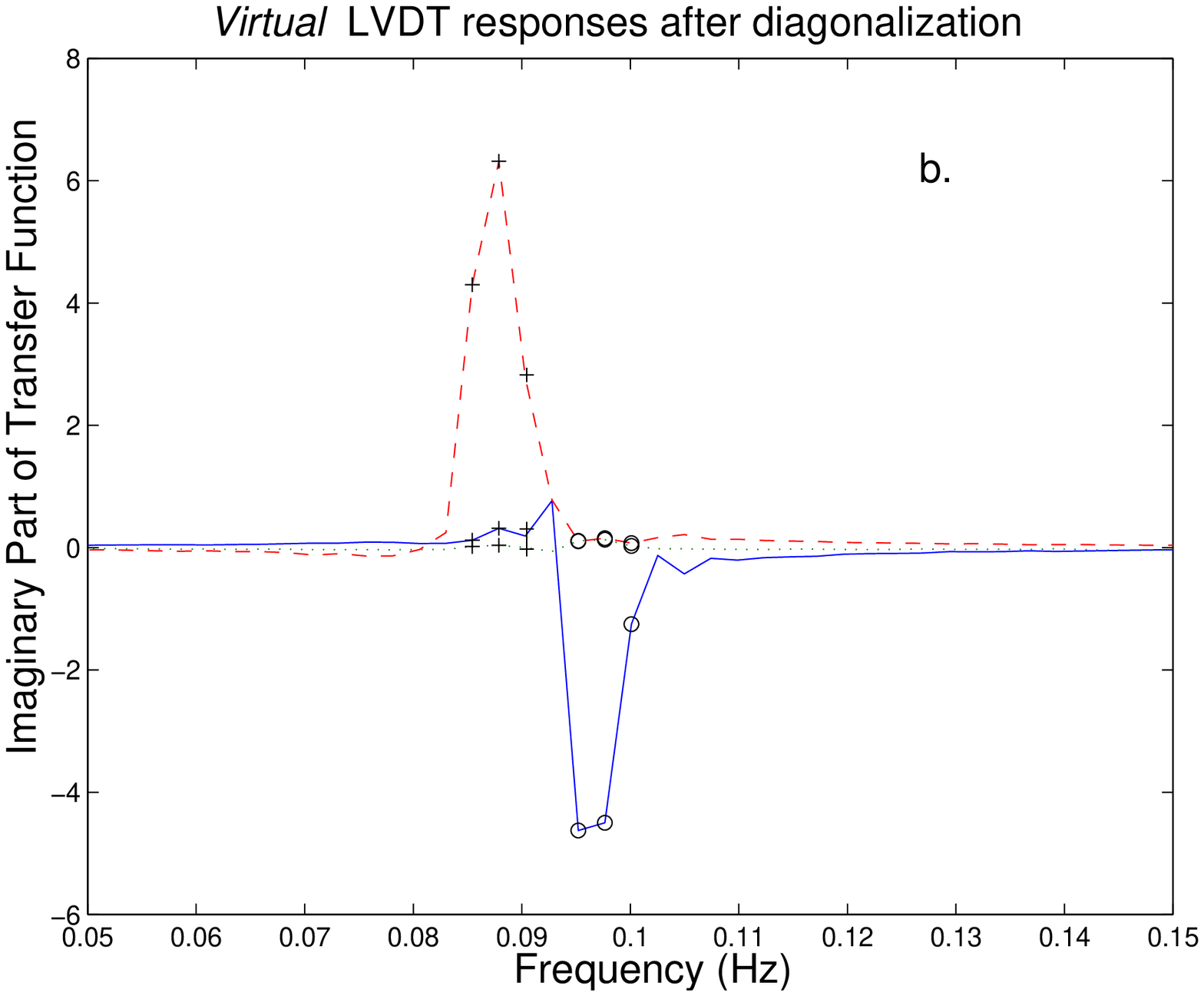}
\end{minipage}
 \caption{\footnotesize Imaginary part of LVDT transfer functions
before (LEFT) and after (RIGHT) the diagonalization procedure. The
{\small +} and {\small o} marks correspond to the integration
points for the $X$ and $Y$ translation modes respectively. The
solid, dashed, and dotted curves in (LEFT) correspond to the
'real' LVDTs whereas in (RIGHT) they correspond to the X, Y and
rotational combinations. Only one of the three coils was energized
for this data set.}\label{figleh} \enc\end{figure}

This merit function is then used with a MATLAB routine FMINSEARCH and the
parameters are varied to find the best result. The sensing matrix obtained by this
so-called 'automatic' diagonalization agrees well with that obtained by other
procedures. We illustrate this by showing some data taken from the injection bench
suspension tower. In fig. \ref{figleh} the response of {\em real} and {\em virtual}
sensors is compared.

In the case where the two IP translational peaks are degenerate, or almost
degenerate, there are two alternative paths.  One is to declare victory and say
that if the modes are degenerate any two orthogonal directions are as good as any
other; then choose the most convenient set.   The other path is to do the best you
can and then look at the very low frequency part of the LVDT spectra and the
accelerometer spectra above several Hertz where most of the resonance activity is
absent.  Simple average values in these regions can be used as merit functions. A
check on the reliability of the results is that the diagonalized modes of the
LVDT's, the accelerometers, and the coil drivers should be approximately related to
each other by known rotations.

\subsection{Selection of the Parameter Set}
The ultimate goal in the diagonalization process is to find a sensing matrix that
transforms sensor outputs into $x-y-\theta$ displacement values and to find a
driving matrix that transforms desired $x-y-\theta$ displacements into driver-coil
currents (see fig. \ref{decomp}). The chosen parameter set should make this task as
simple as possible and should have a clear association with the component systems.

Certain parameters are fixed and known: the positions and orientations of the
sensors and coils. Other parameters such as the relative sensitivity of the sensor
and drive-coil components should be within a few percent of unity.  Finally, if
there is no mixing between the two translational modes and the rotational mode,
there is only one parameter that we have no a priori knowledge: the angle $\psi$
between the translational mode coordinate system and the laboratory system.  We
start our discussion with the simplest case; all system component sensitivities are
the same and, also, there is no mode mixing. We will add these complications later.

The components of the horizontal LVDT's, accelerometers and coil driver systems
are arranged in groups of three, positioned 120 degrees apart with respect to each
other. We denote a system state by a vector $s =[ a\ b\ c]$, the values of the
three components. A translation and/or rotation of the IP results in signals in the
sensor components given by $$s = \left( \bega{c} a \\ b \\ c \endar\right) = \bf D
\left( \bega{c} u \\ v \\ r\ \theta \endar\right)$$ Here, $u$ and $v$ are
translations, $\theta$ is a rotation and $r$ is the radius of the sensor system.
The  $3\times 3$ matrix $\bf D$ contains the geometry of the sensor system. For
example given a set of 3 sensors at respective $u-v$ angles of $\phi,\ \ \phi+120\
{\rm and}\ \phi+240$ degrees: the $\bf D$ matrix is:

$${\bf D} = \left[ \bega{ccc} \sin(\phi) & \cos(\phi) & 1\\
 \sin(\phi+2\pi/3) & \cos(\phi + 2\pi/3) & 1\\
 \sin(\phi+4\pi/3) &  \cos(\phi+4\pi/3) & 1   \endar\right]$$
The relative gain of the sensors can be easily incorporated at this point  by
letting $\bf{D}$ be multiplied by a  $3\times 3$ diagonal gain matrix: $  {\bf
D\Longrightarrow G\ D}$

A 3-vector $s$ describing the state of one of the system can be transformed back
into its equivalent spatial representation; a $u-v$ translational vector plus a
rotation.
 This is obtained by multiplying the state vector
by a $3\times 3$ matrix, $\bf R$: $$ \left( \bega{c}u \\ v \\ r\ \theta\endar
\right) = \bf{R}\ \left( \bega{c} a \\ b \\ c \endar\right) \hskip .3in {\rm
where}\ \bf R = \bf D^{-1}$$
\medskip

Thus, for example, given $\phi=0$ , $\bf D$ and $\bf R$ become

$${\bf D} = \left[ \bega{ccc} 0 & 1 & 1\\ -0.866 & -0.5 & 1\\ 0.866 & -0.5 & 1
\endar\right] \hskip .4in \bf R = \left[ \bega{ccc} 0 & -0.577 & 0.577 \\ 0.667 &
-0.333 & -0.333 \\ 1/3 & 1/3 & 1/3   \endar\right]  $$ Multiplying the first two
rows of $\bf R$ by the scaling factor $\sqrt{3/2}$ and the last row by  $\sqrt{3}$
we obtain the 'standard' sensing matrix (where each line is normalized to 1):
$$ \bf S = \left[ \bega{ccc} 0 & -0.707 &
0.707\\ 0.817 & -0.408 & -0.408\\ 0.577 & 0.577 & 0.577  \endar\right]$$

The sensing matrix $\bf S$  happens to be an orthogonal matrix corresponding to a
rotation in three dimensions. This rotation can be constructed by choosing three
Euler angles according to the following prescription:
\begin{enumerate}
\item Rotate an amount $\theta = 45^\circ$ clockwise about the $z$-axis.
\item Rotate an amount $\phi = \cos^{-1}(1/\sqrt 3) = 54.7^\circ $
  clockwise about
the  $y^\prime$-axis.
\item Rotate an amount $\psi = 210-\phi_0$ clockwise about the
the $z^{\prime\prime}$ axis. Here, $\phi_0$ is the angle of the sensor system.
\end{enumerate}
We note that the values for $\theta$ and $\phi$ do not depend on the angle of the
sensor system nor on the direction of the $x-y$ axes. Thus for this most {\it
simple case}, equal component gains  and no mode mixing, a sensing matrix can be
constructed by finding the correct relative angle between the sensor system and the
$x-y$ axes. A scan in the variable $\psi$ is all that is needed.

\bigskip{\bf Complications}

\noindent The case where the sensor components have different relative gain has
been discussed; we multiply $\bf D$ by a diagonal gain matrix which gives ${\bf
D\Longrightarrow G\ D}$ and $\bf S \Longrightarrow
 \bf D^{-1} \bf G^{-1}$.

If the relative angles of the sensor components are not exactly $120^\circ$ apart
the resulting $\bf S$ matrix can no longer be described by an orthogonal rotation
but must be slightly modified.  Suppose, for example, in the relation $s = (a\ b\
c) = {\bf D} (x\ y\ r\theta)$ the element corresponding to $a$ is rotated by a
small angle $\gamma$.  Then a new vector
 $$s^\prime = \left( \bega{c} a^\prime \\ b \\ c \endar\right)
= \bf A  \left( \bega{c} a \\ b \\ c \endar\right)= \left[ \bega{ccc} \alpha &
\beta  & -\beta \\ 0 & 1& 0\\ 0& 0& 1 \endar\right] \left( \bega{c} a \\ b \\ c
\endar\right)$$ can be defined with $ \alpha = \cos (\gamma) $ and $\beta = \sin
(\gamma)/\sqrt{3}$. The modified sensing matrix becomes \\ $\bf S \Longrightarrow
 \bf D^{-1} \ \bf A^{-1}$.   The effect is not large, a misalignment of 10 mrad
mixes the components by one percent or so.

In summary we have identified five parameters to define the sensing matrix; one
angle $\psi$, two relative sensor gains, $g2/g1$, $g3/g1$, and two misalignment
angles $\gamma$ and $\rho$.  This parameter space is then searched to maximize the
merit function. In practice, we have found that it is sufficient to vary the single
parameter $\psi$ to obtain a set of diagonalization parameters enabling us to close
the inertial damping feedback loops. Fig. \ref{xth} shows the effect of the
diagonalization on the SA transfer function.

\section{Inertial damping: principle}

The control we describe here is called {\it inertial damping} because it is
performed by using (mostly) {\it inertial sensors} (accelerometers).
In the following, with the help of a simple model, we explain why this
is the best choice to achieve a high performance damping.

Let us consider a simple pendulum of mass $m$ and length $l$. Let $x$ be the
abscissa of the suspended mass, $x_0$ that of the suspension point. Let $F_{\rm
fb}$ the external force on the pendulum (i.e. the feedback force to control it).
The equation of motion is then:
\beq F_{\rm fb}=m\ddot{x}+\gamma\dot{x}+k(x-x_0) \label{eqschema} \eeq
where $\gamma$ is the viscous dissipation factor and $k=mg/l$. The control loop of
such a system is sketched in fig. \ref{schema}, where $H(s)$ is the mechanical
transfer function, $G(s)$ is the compensator and {\it out} is the output of the
sensor used. The goal of the control is the damping of the pendulum resonance. This can be
done easily with a {\it viscous} (theoretical) feedback force:
\beq F_{\rm fb}=-\gamma'\dot{x} \eeq
Our sensors do not measure $x$. Their output is:
\beq {\mathit out}=\left\{\begin{array}{ll} x-x_0 &\;\; \mbox{for displacement
sensors} \\ \ddot{x} &\;\; \mbox{for accelerometers}
\end{array} \right. \eeq
Therefore, the actual ``viscous" force that can be built if position sensors are
used has the form:
\beq F^p_{\rm fb}=-\gamma'\frac{\rm d}{{\rm d}t}(x-x_0) \eeq
It can be easily shown that with such a feedback force the closed loop equation of
motion (in Laplace space) reduces to:
\beq x(s)=\frac{\omega_0^2+G_0s}{s^2+\omega_0^2+(\omega_0/Q+G_0)s}\cdot x_0(s)
\label{pcltf} \eeq
where $G_0=\gamma'/m$ is a gain parameter (in a real feedback system a
frequency dependent gain function $G(s)$ rather than a fixed gain parameter has to
be considered) measuring the intensity of the viscous feedback force, and $Q$ is
the open loop quality factor. When the loop is closed a damping of the resonance is
achieved:
\beq Q'\stackrel{G>>1}{\longrightarrow}\frac{\omega_0}{G} \eeq
Nevertheless, as the gain is increased, a larger amount of noise is reinjected
off-resonance. This is associated to the term ``$G_0s$" in the numerator of
(\ref{pcltf}) and depends on the fact that the sensor used to build up the feedback
force measures the position of the pendulum with respect to ground. Therefore, an
infinitely efficient feedback would ``freeze" the pendulum to ground (which is
seismic noisy), reducing its motion at the resonance, with the drawback of
bypassing its attenuation properties above resonance.

The situation is fairly different when an inertial sensor is used. In this case the
viscous feedback force is obtained by integrating the accelerometer output, which
does not depend on $x_0$:
\beq F^a_{\rm fb}=-\gamma'\int \ddot{x}{\rm d}t \eeq
The closed loop equation of motion is then:
\beq x(s)=\frac{\omega_0^2}{s^2+\omega_0^2+(\omega_0/Q+G_0)s}\cdot x_0(s)
\label{acltf} \eeq
A damping of the resonance is obtained (exactly as in the previous case) but
without reinjection of off-resonance noise. In fig. \ref{apcomp} a simulation of
the closed loop transfer function $x(s)/x_0(s)$ is shown in the two cases.

Up to now we have considered a simple viscous damping. It is possible to increase
the bandwidth of the control if the feedback force contains a term proportional to
$x$ (the double integral of the accelerometer signal). The result obtained in this
case is shown in fig. \ref{xdamp}.

\section{Control strategy}

In this section we extend the principles of the previous section and describe the
strategy to control the SA.

The basic idea of inertial damping is to use the accelerometer signal to build up
the feedback force. As a matter of fact a perfect feedback using only the inertial
sensor information would null the acceleration of the pendulum but it would do
nothing if the pendulum moved at constant velocity: such a control is unstable with
respect to drifts. Therefore, if the control band is to be extended down to
DC, a position signal is necessary. Our solution is to merge the two sensors: the
virtual LVDT (position) and accelerometer signals are combined in such a way that
the LVDT signal ($l(s)$) dominates below a chosen crossover frequency $f_{\rm merge}$
while the accelerometer signal ($a(s)$) dominates above it (see fig. \ref{merge}).
  The feedback force has the form :
\beq F_{\rm fb}=G(s)\left[a(s)+\epsilon l(s)\right] \eeq
where $G(s)$ is the transfer function of the compensator and
$\epsilon$ is the parameter whose value determines $f_{\rm merge}$. We choose
$f_{\rm merge}\sim 10$ mHz (corresponding to $\epsilon\sim 5\cdot 10^{-3}$). This
approach stabilizes the system with respect to low frequency drifts at the cost of
reinjecting a fraction $\epsilon$ of the seismic noise via the feedback. In order
to reduce the amount of reinjected noise at $f>f_{\rm merge}$ (while preserving
feedback stability at the crossover frequency) the LVDT
signal $l(s)$ is properly low-pass filtered.

We describe in the following the feedback design for the three d.o.f., starting
from the the translational ones. The virtual $X$ and $Y$ sensors show many resonant
peaks (the modes of a chain of pendulums) and this requires a more sophisticated
feedback strategy. The digital filter used to control the translation modes
($G(s)$) is shown in fig. \ref{digflt} (LEFT). It shows three main features:
\begin{itemize}
\item for $0.01<f<2$ Hz the gain is proportional to $f^{-2}$. This
corresponds to the case of fig. \ref{xdamp}: the accelerometer signal is integrated
twice and the feedback force is proportional to $x$;
\item for $f>2$ Hz the gain is proportional to $f^{-1}$. The
accelerometer signal is integrated once: the feedback force is proportional to the
velocity and a viscous damping is achieved;
\item the peaks visible in the filter are necessary to compensate
the corresponding dips in the mechanical transfer function ($H(s)$) of fig.
\ref{xth}, in order to make the feedback stable.
\end{itemize}
Fig. \ref{digflt} (RIGHT) shows the open loop gain transfer function $G(s)H(s)$.

The damping strategy for the $\Theta$ mode is simpler: the $\Theta$ virtual sensor
shows one resonance peak only and no dips (see fig. \ref{xth}, RIGHT): no compensation
is necessary. Apart from this, the feedback strategy is similar to the ones used
for the translational modes.
\section{Inertial damping: experimental results}

Results of the inertial control (on three d.o.f.) are shown in figure
\ref{results}. This measurement was performed in air on the prototype SA.
The noise at the top of the
IP is reduced over a wide band (10 mHz - 4 Hz). A gain greater than 1000 was
obtained at the main SA resonance (0.3 Hz). The RMS translational motion of the IP
(calculated as $x_{\rm RMS}(f)=\sqrt{\int_f^\infty \tilde{x}^2(\nu){\rm d}\nu}$) in
10 sec. is reduced from more than 30 to 0.3 $\mu$m. The closed loop floor noise
corresponds to the fraction of seismic noise reinjected by using the position
sensors for the DC control and can, in principle, be reduced by a steeper low pass
filtering of the LVDT signal at $f>f_{\rm merge}$ and by lowering $f_{\rm merge}$:
both this solutions have drawbacks and require careful implementation.

The control strategy adopted in this experiment requires a careful compensation of
the dips in the transfer function. A less aggressive strategy has been adopted for
the SA on VIRGO site. The filter is a simple integrator (plus compensation of high
frequency structural resonance). The gain is lower but no compensation of the dips
is needed. This makes the control loops more robust with respect to changes of the
frequency of the poles and zeroes in the transfer function that might be induced by
temperature variations.
Results (obtained under vacuum, on the VIRGO site) are shown in fig.
\ref{virgodamp} for the 3 d.o.f.. Even if the gain of the loop is less in this
case, the measured closed loop noise floor is lower because the system is under
vacuum.

{\em Inertial damping} is a technique for damping the normal modes of the VIRGO
suspension and reducing seismic noise entering the system. The measurements
presented in this paper demonstrate the success of the method. Further measurements
of the mirror motion with and without inertial damping are necessary to prove
directly the reduction of the residual mirror motion. These measurements will be
the topic of a forthcoming paper.

\pagebreak

\newpage

\bef \bec
\begin{minipage}{0.4\textwidth}
    \includegraphics[bb=2.5cm 1.5cm 17cm
28cm,clip=true,width=\textwidth]{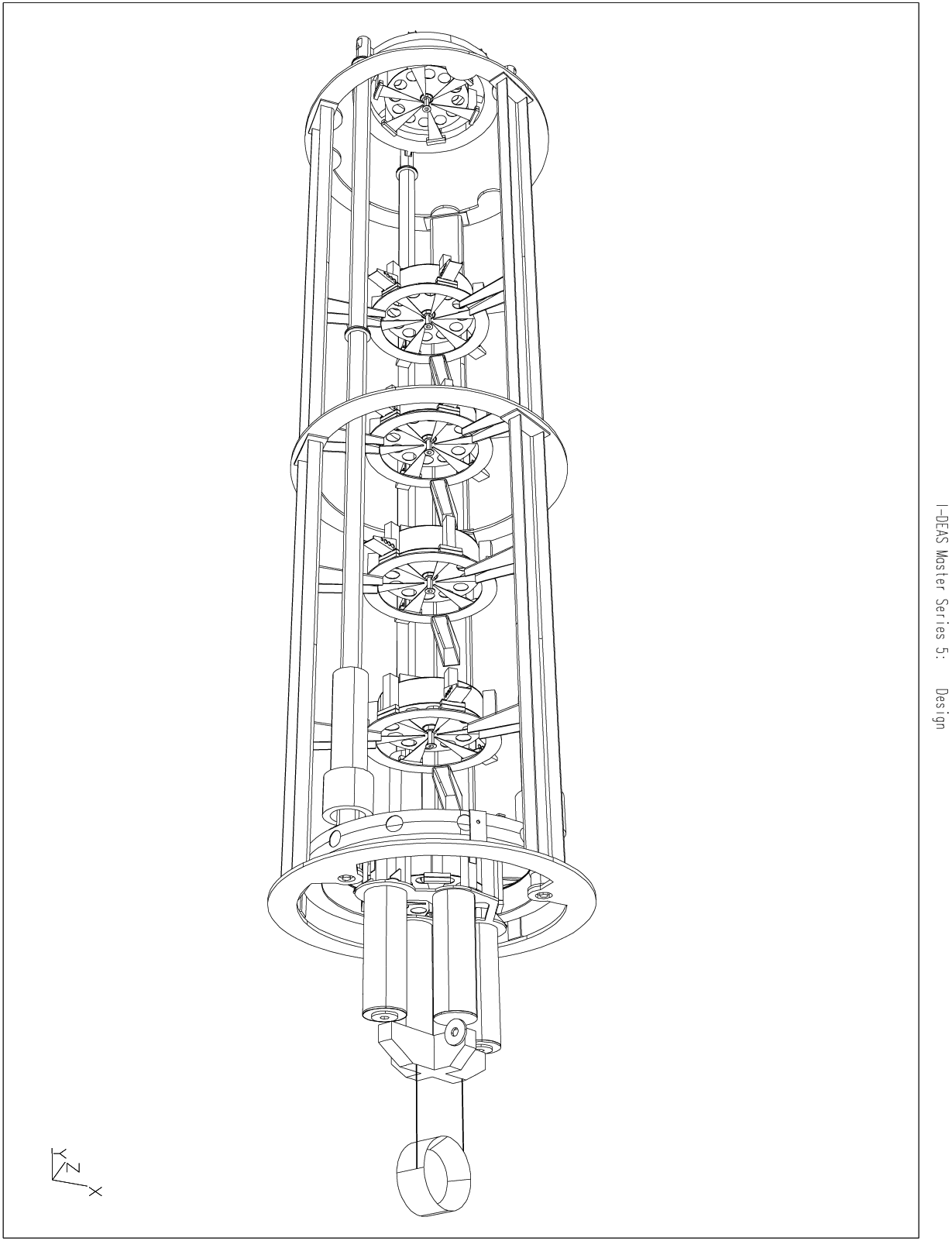}
\end{minipage}\hfill
\begin{minipage}{0.6\textwidth}
 \includegraphics[width=.9\textwidth]{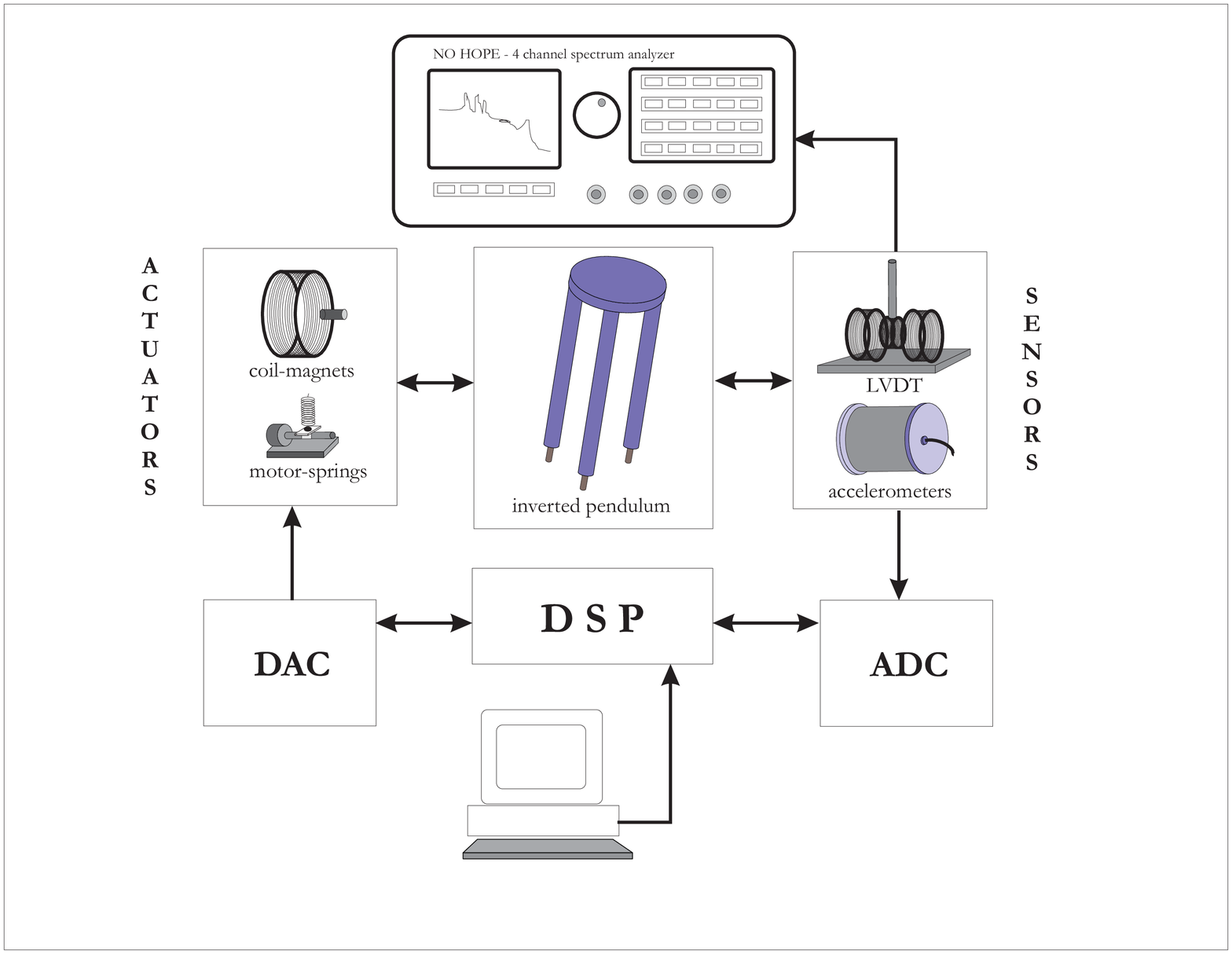}
 \includegraphics[width=0.95\textwidth]{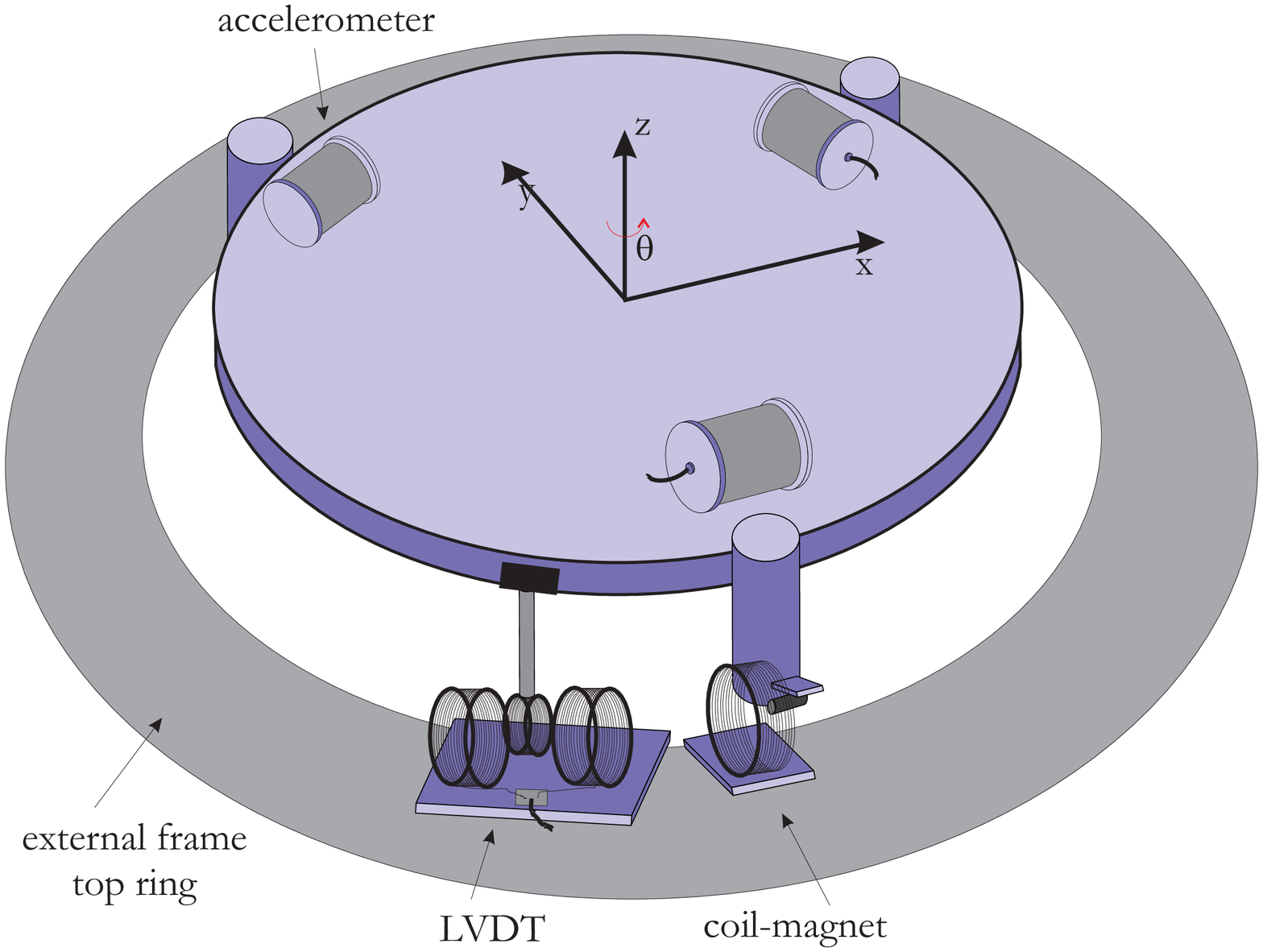}
\end{minipage}
\caption{\footnotesize LEFT: the superattenuator; RIGHT TOP:
logical scheme of the setup for the local active control; RIGHT
BOTTOM: simplified view of the IP top table, provided with the 3
accelerometers. One LVDT position sensors and one coil-magnet
actuator are also shown. } \label{setup} \enc\enf

\bef \bec \ingr[width=10cm]{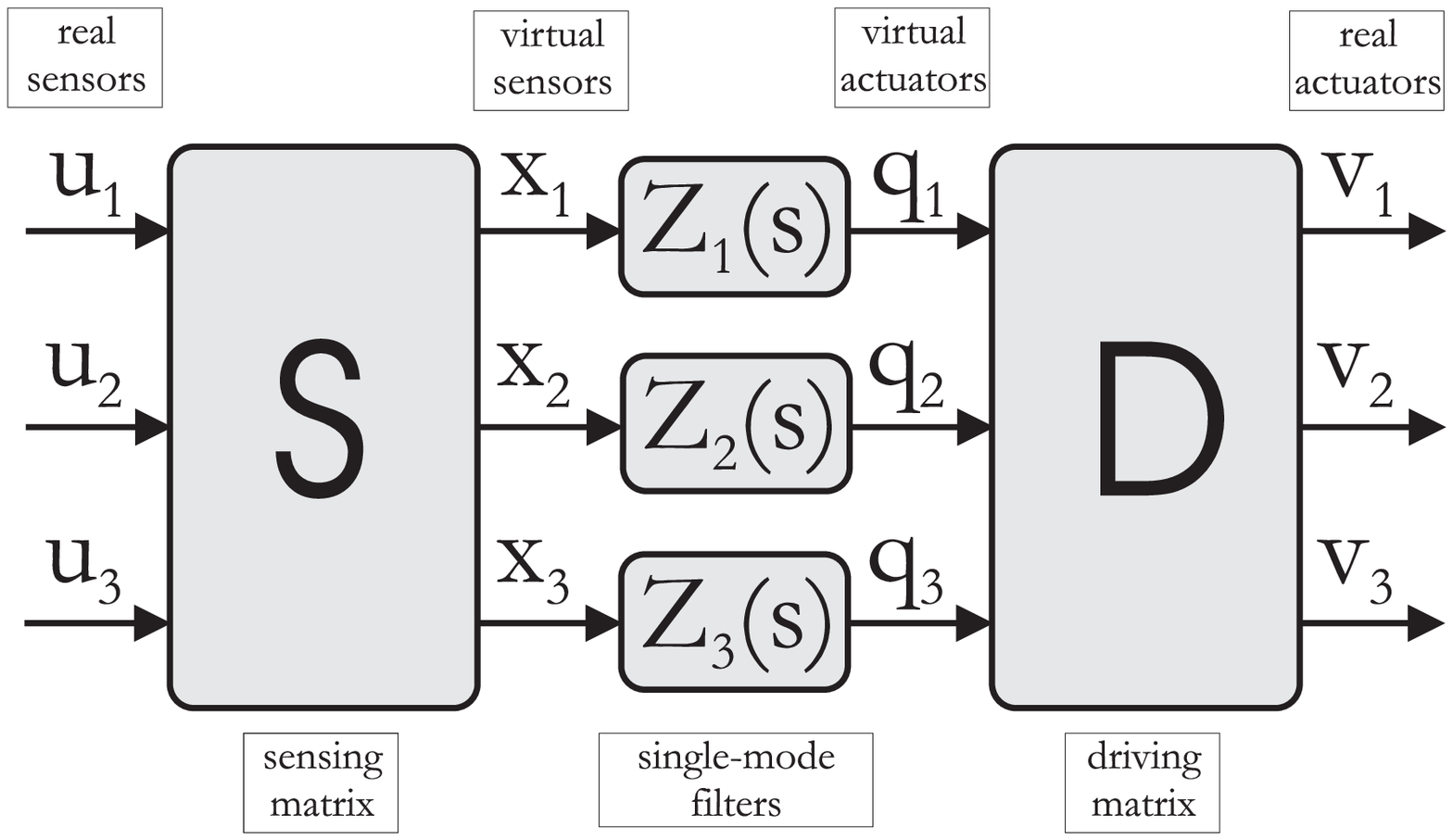} \caption{\footnotesize
Decomposition of a 3-modes system into 3 non-interacting 1-mode
systems: the real sensors signals $x_i$ are recombined by the
matrix $S$ to create the virtual sensors $x_i$. Each virtual
sensor is acted upon independently. The virtual forces $q_i$ are
defined by the filters $z_i$. The virtual forces are converted
into voltages driving the real actuators by the matrix $D$.}
\label{decomp} \enc\enf

\bef\bec
\begin{minipage}{.47\textwidth}
\ingr[width=.95\textwidth]{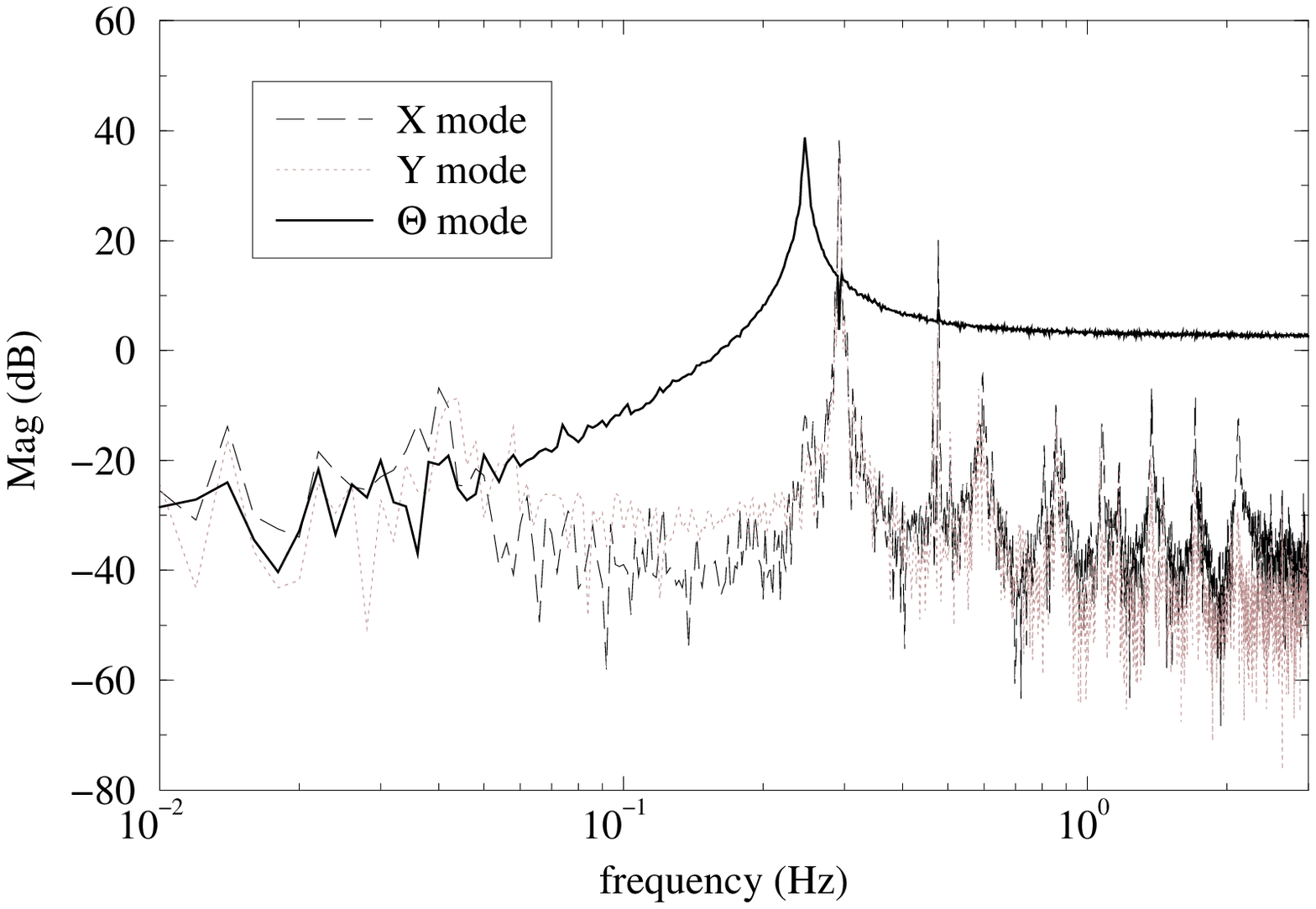}
\end{minipage}\hfill
\begin{minipage}{.47\textwidth}
    \includegraphics[width=\textwidth]{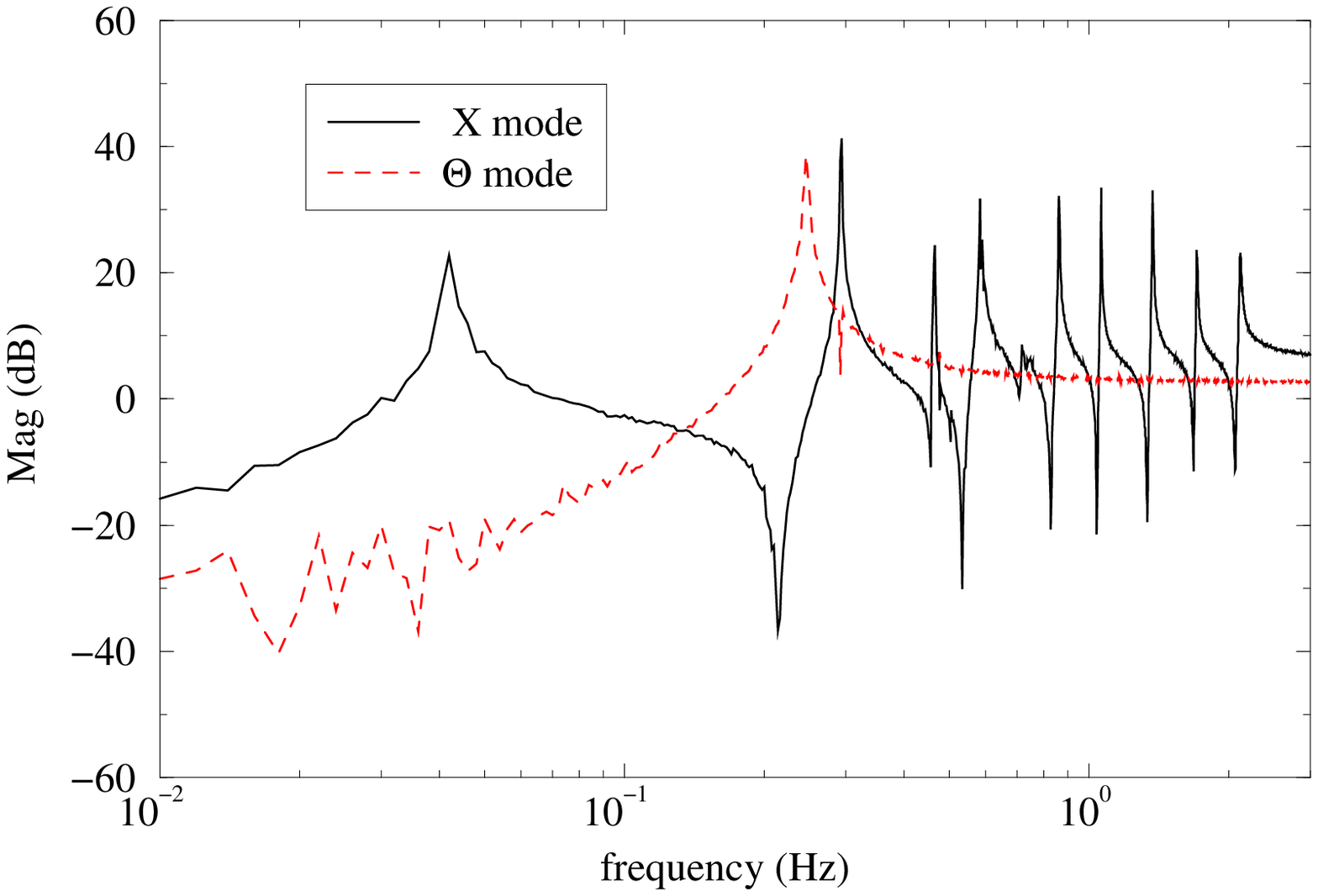}
\end{minipage}
\caption{\footnotesize Effect of the digital diagonalisation.
LEFT: the output of the 3 virtual accelerometers when $\Theta$ is
excited. RIGHT: the output of the virtual accelerometers
    $X$ and $\Theta$ are compared. Different feedback strategies are
    required in the two cases, because $X$ senses all the translational
    modes of the SA chain. }\label{xth}
   \enc  \enf

\bef\bec \ingr[width=.8\textwidth]{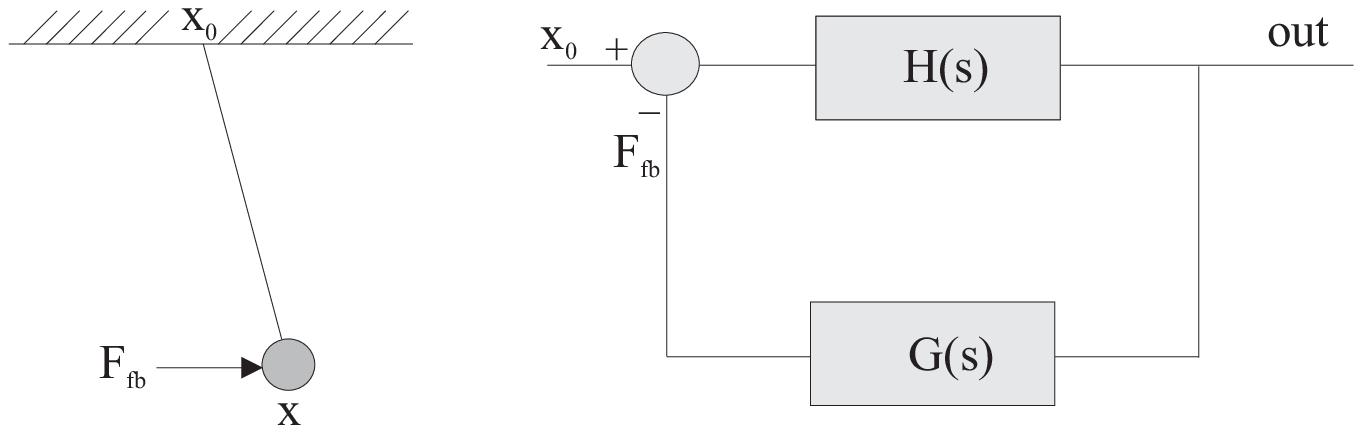}
 \caption{\footnotesize The control scheme for a simple
pendulum: $H(s)$ is the mechanical transfer function, $G(s)$ the
compensator filter.} \label{schema}\enc\enf

\bef\bec
\begin{minipage}{.47\textwidth}
\ingr[width=\textwidth]{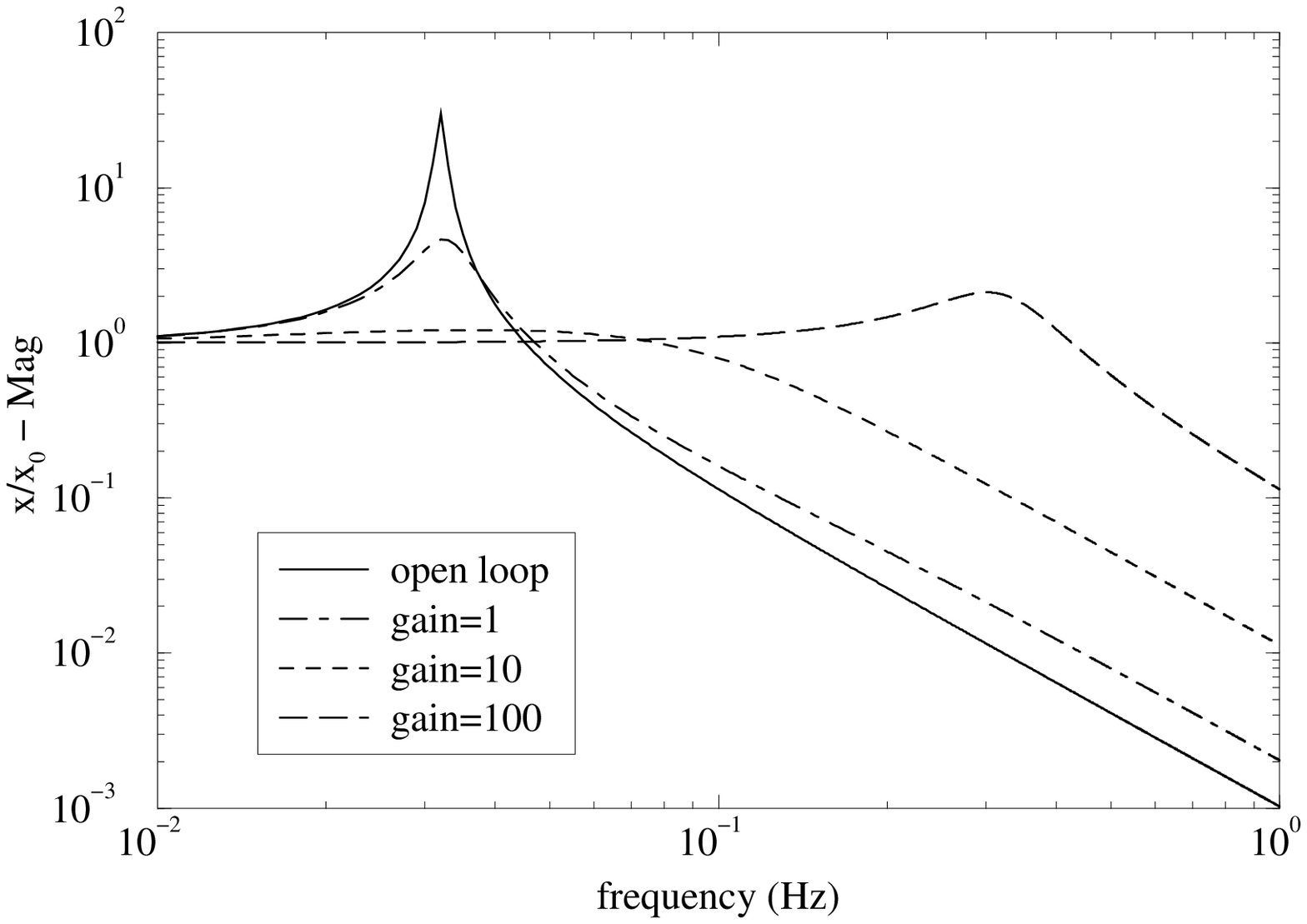}
\end{minipage}\hfill
\begin{minipage}{.47\textwidth}
\ingr[width=\textwidth]{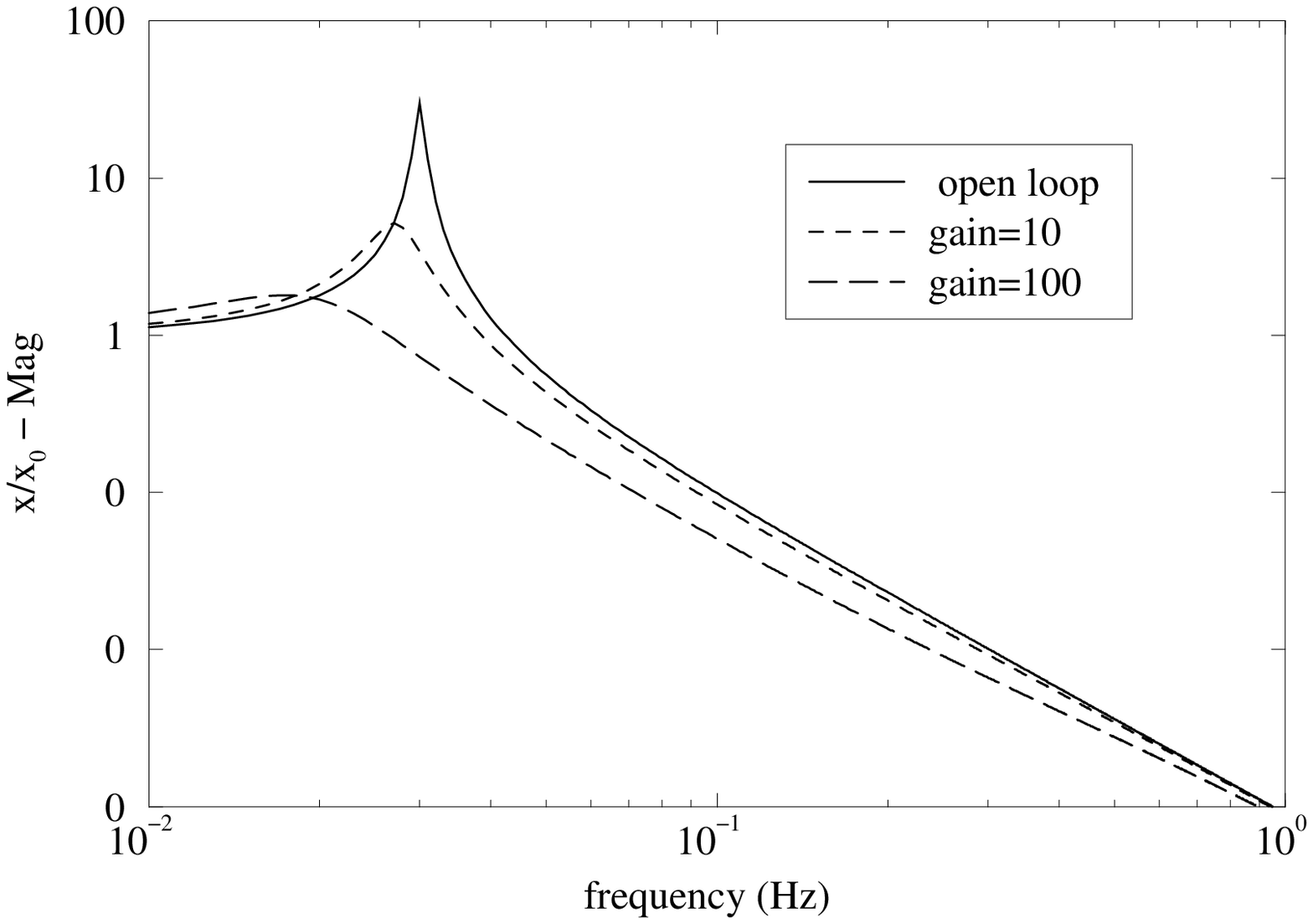}
\end{minipage}
\caption{\footnotesize Damping of a simple pendulum: the closed
loop transfer function $x(s)/x_0(s)$ (magnitude) when a position
sensor is used (LEFT) and when an accelerometer is used (RIGHT).
In the first case it is evident the re-injection of noise above
the resonance.} \label{apcomp} \enc\enf

 \bef\bec
\includegraphics[width=0.47\textwidth]{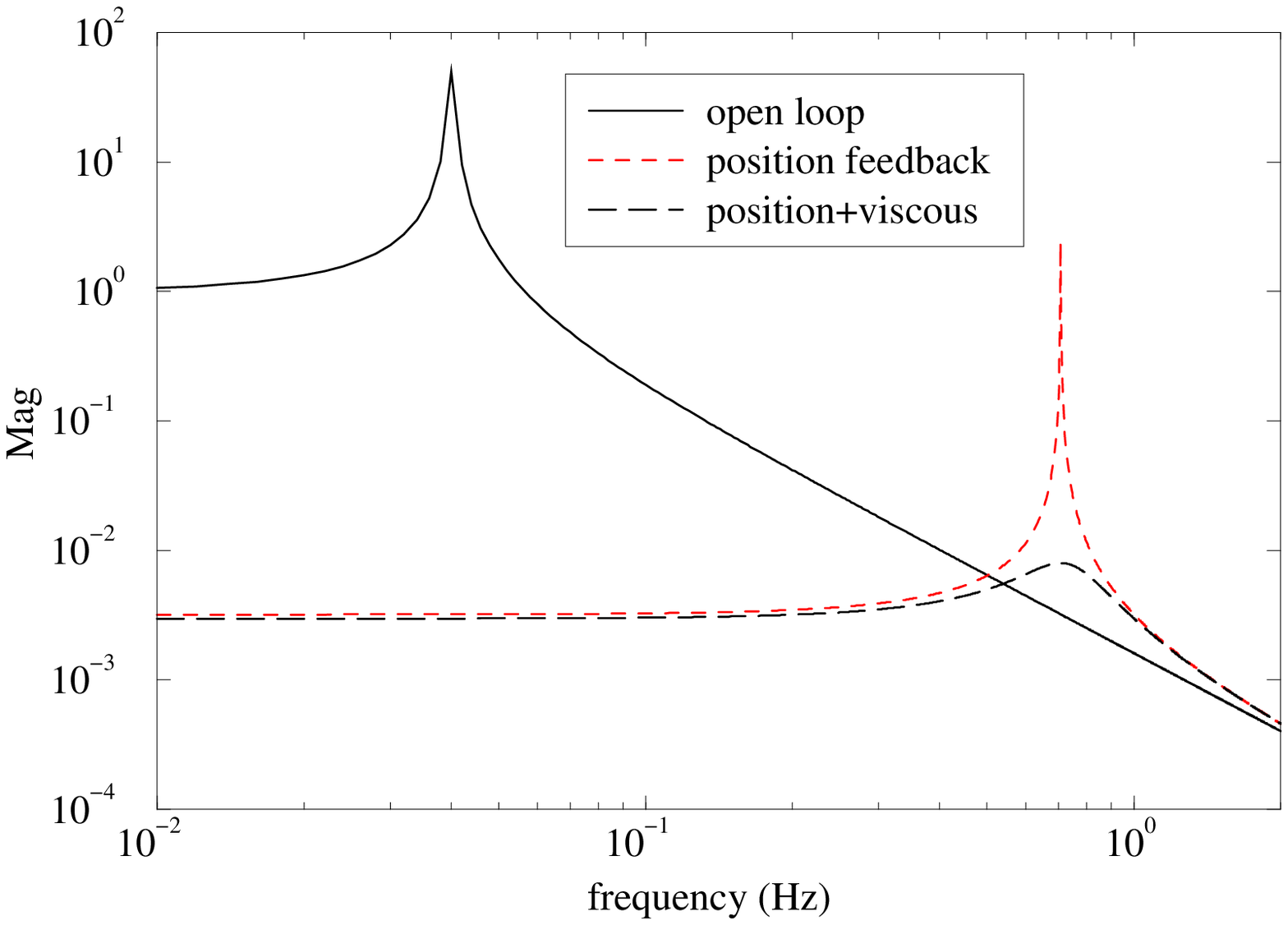}
 \caption{\footnotesize Inertial damping of a simple
pendulum when a position feedback is implemented. } \label{xdamp}
\enc\enf

\bef\bec
\includegraphics[width=0.47\textwidth]{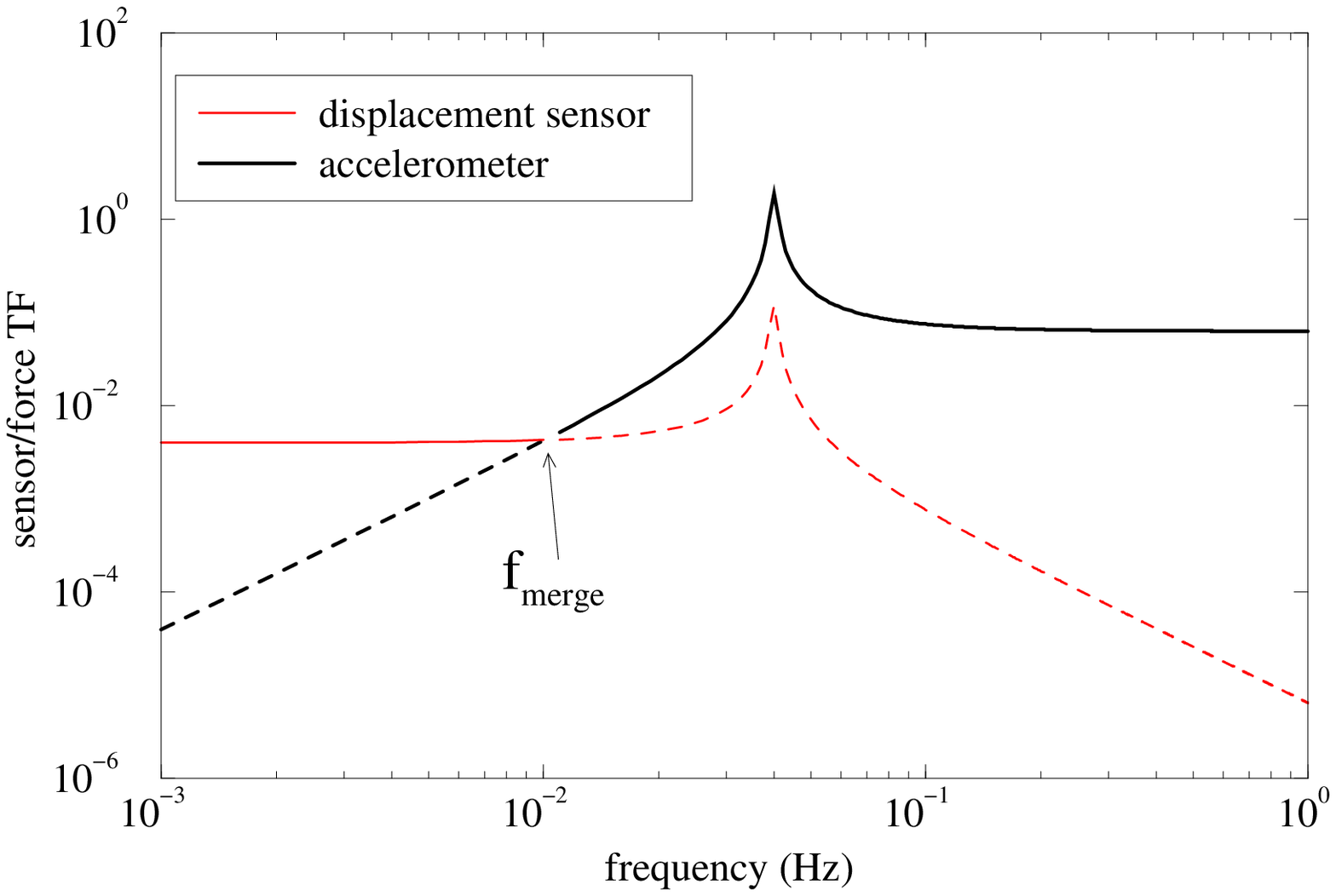}
\caption{\footnotesize {\it Merging} of displacement and
acceleration sensors (simulation for a simple pendulum).}
\label{merge} \enc\enf
\newpage
\bef\bec
\begin{minipage}{.47\textwidth}
\ingr[bb=130 60 693 473,clip=true,width=\textwidth]{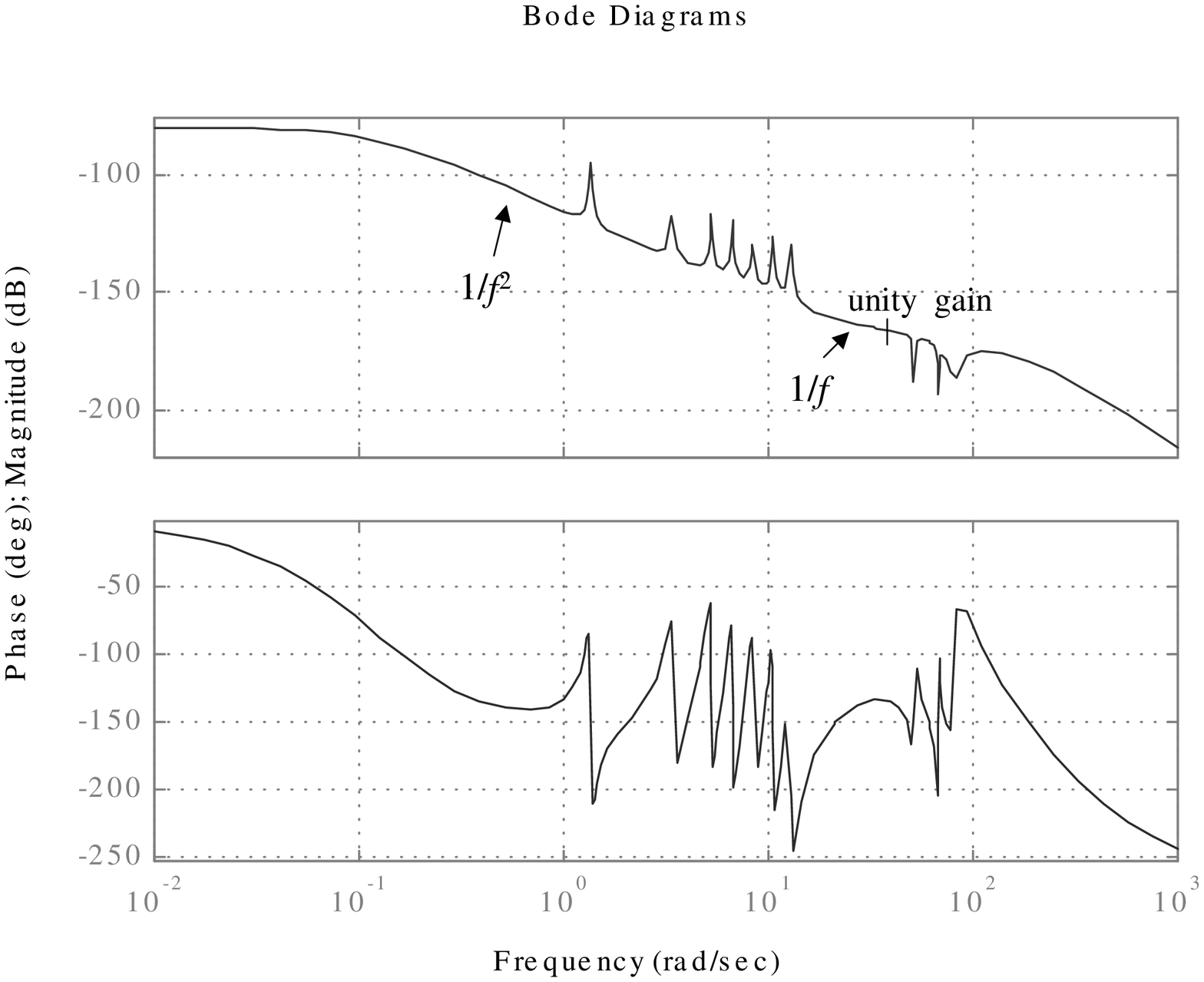}
\end{minipage}\hfill
\begin{minipage}{.47\textwidth}
\ingr[width=\textwidth]{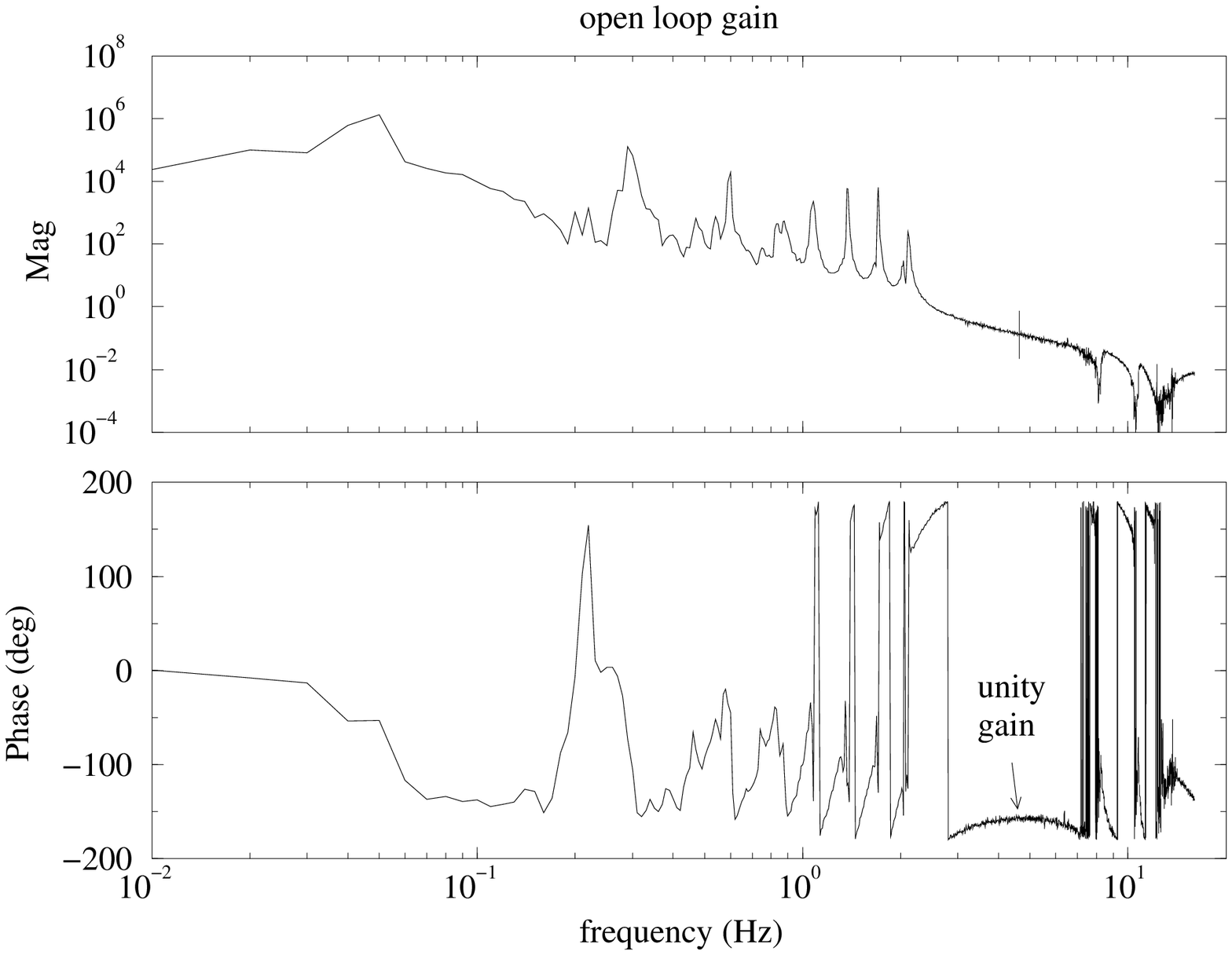}
\end{minipage}
\caption{\footnotesize LEFT: Digital filter used for the inertial
damping of a translation mode ($X$). The filter slope is $f^{-2}$
in the range 10 mHz$<f<$3 Hz, $f^{-1}$ for $f>$3 Hz. The unity
gain is at 4 Hz. The peaks in the digital filter are necessary to
compensate the dips in the mechanical transfer function (see the
transfer function of the $X$ mode in fig. \ref{xth}). RIGHT: open
loop gain function (measured). The phase margin at the unity gain
frequency is about $25^\circ$.} \label{digflt} \enc\enf

\bef\bec
\begin{minipage}{.47\textwidth}
\includegraphics[width=\textwidth]{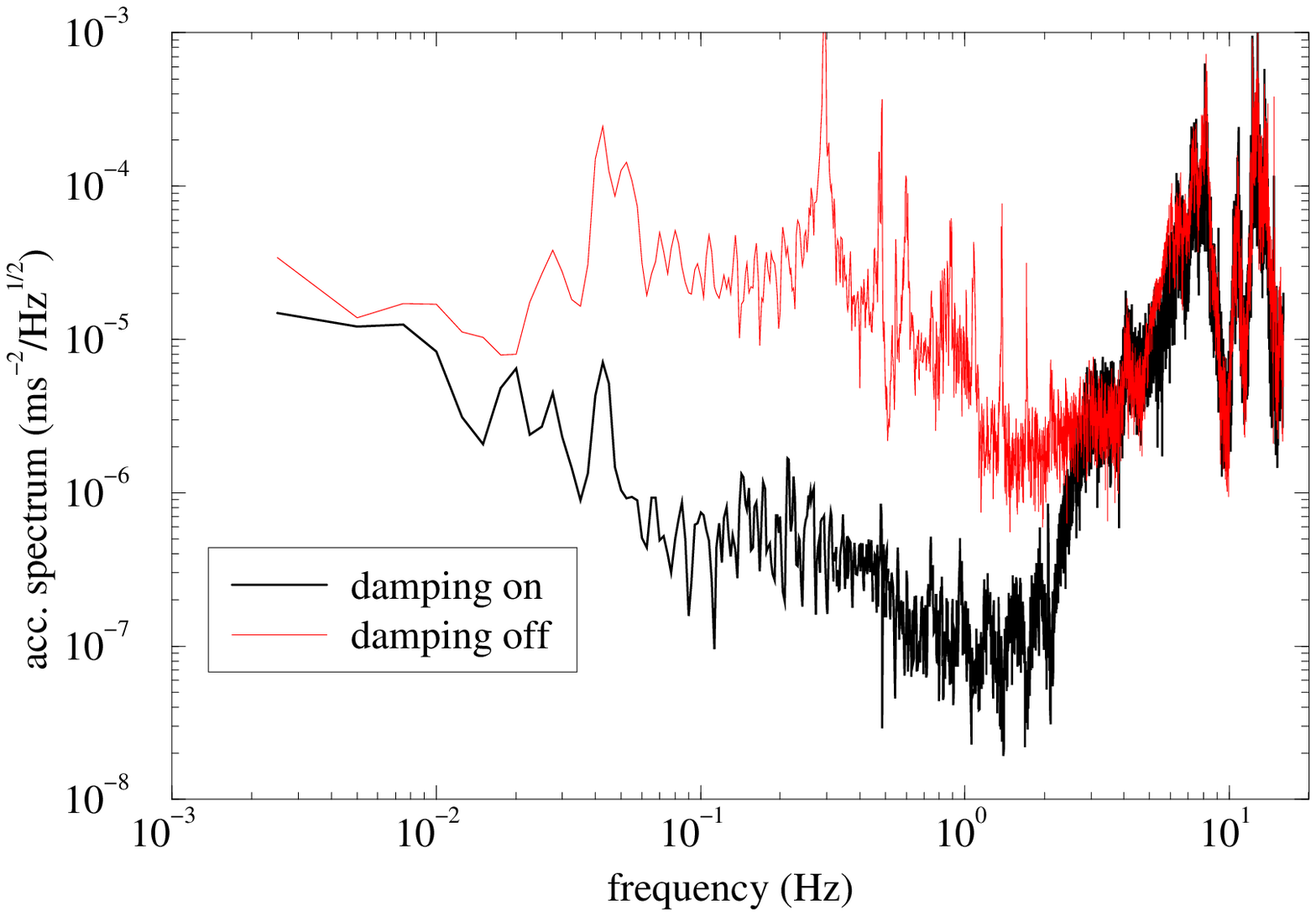}
\end{minipage}\hfill
\begin{minipage}{.47\textwidth}
\includegraphics[width=\textwidth]{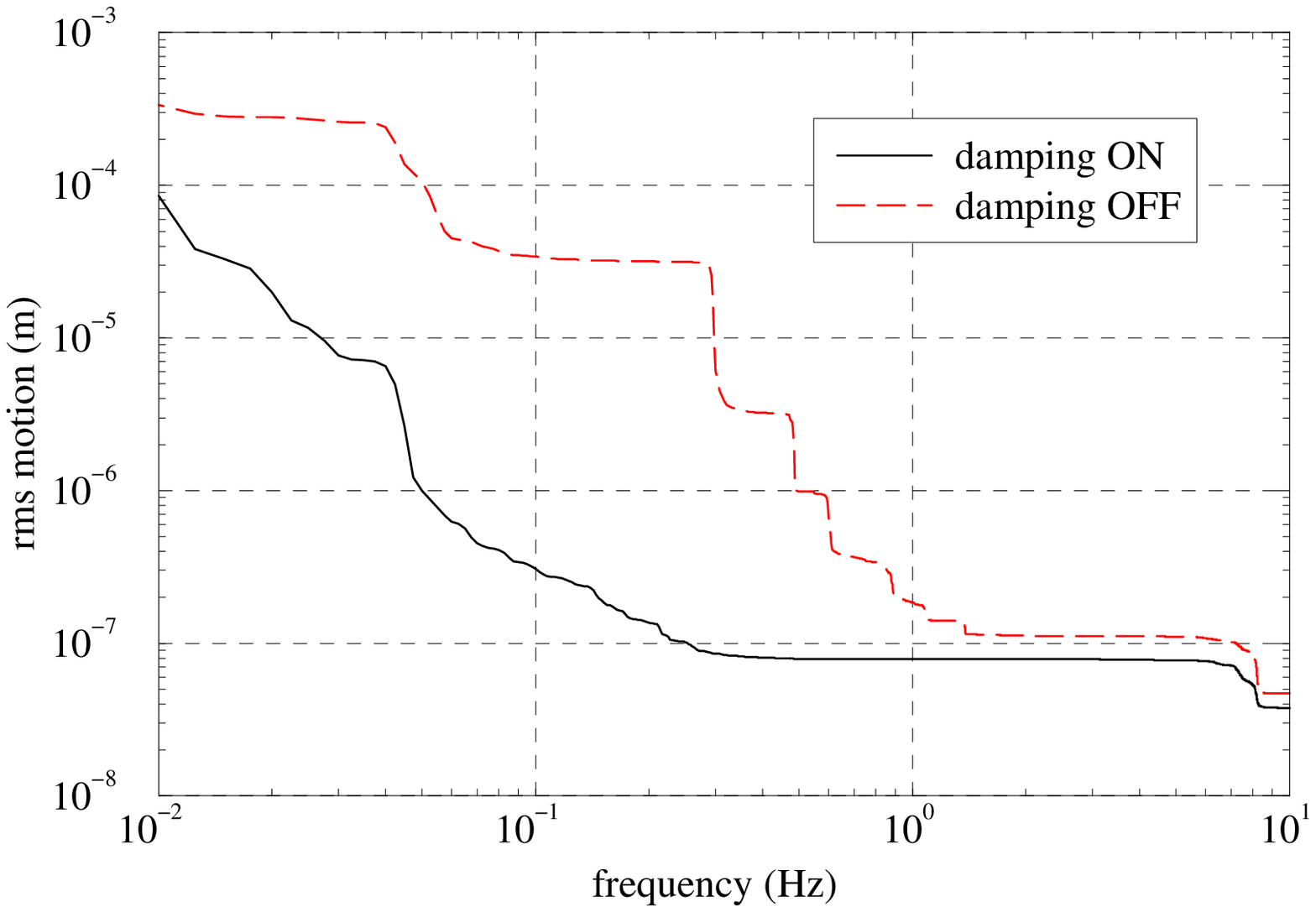}
\end{minipage}
\caption{\footnotesize Performance of the inertial control
($X,Y,\Theta$ loops closed) of the superattuenuator, measured on
the top of the IP: the left plot shows the acceleration spectral
density as measured by the {\it virtual} accelerometer $X$
(translation). The right plot shows the effect of the feedback on
the RMS residual motion of the IP as a function of the frequency.}
\label{results} \enc\enf

\bef\bec
 \ingr[width=.7\textwidth]{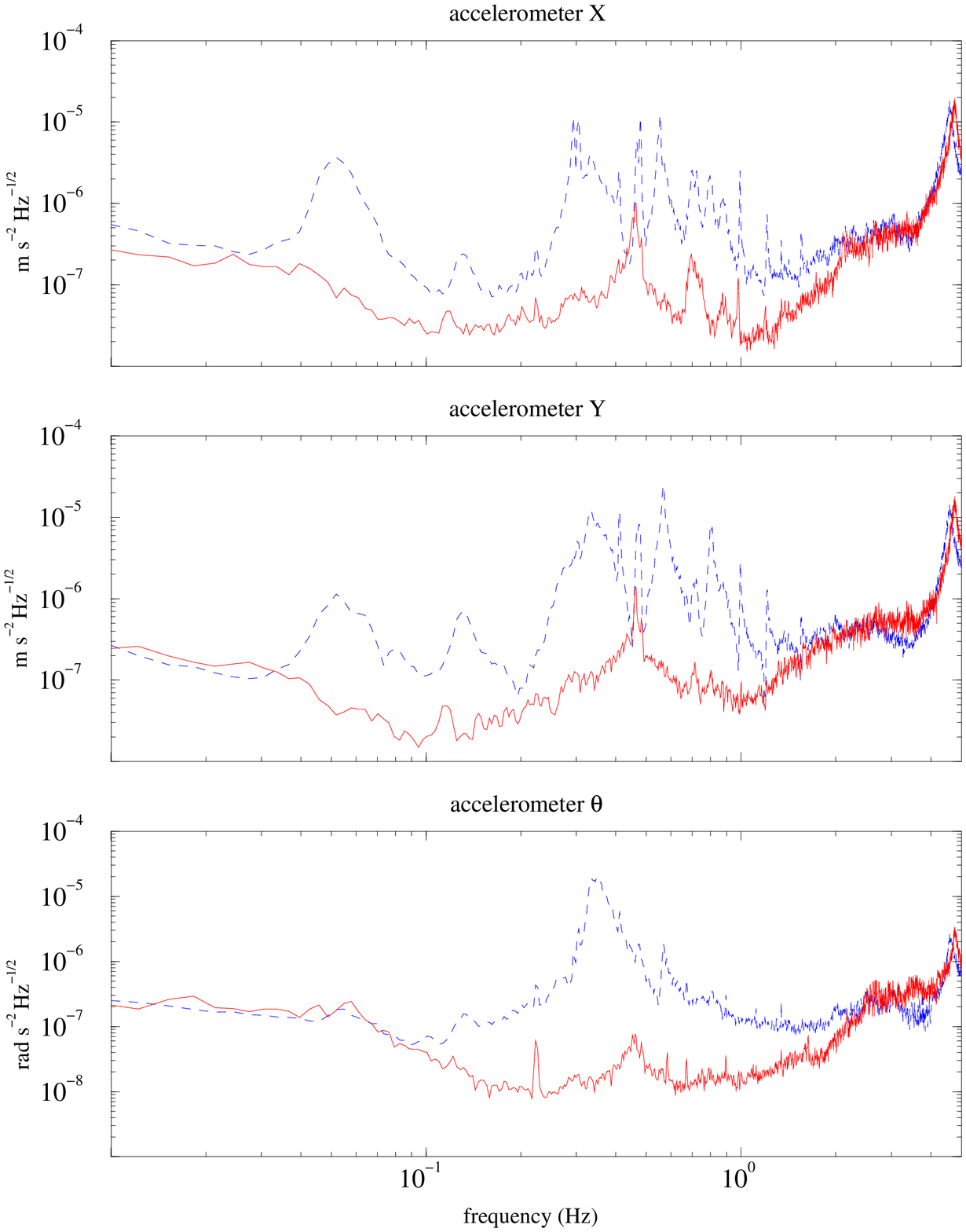}
\caption{\footnotesize Performance of the inertial damping on one
of the VIRGO superattenuators (the 3 d.o.f. $\Theta, Y, X$ are
shown, the system was under vacuum).} \label{virgodamp} \enc\enf

\end{document}